\documentclass[twocolumn,aps,pra,showpacs,floatfix,superscriptaddress,reprint]{revtex4-1}
\usepackage{graphicx}
\usepackage{amsmath}
\usepackage{bm}
\usepackage{color}
\usepackage{dcolumn}

\begin{document}


\newcommand{\cmt}[1]{\textbf{[\![{#1}]\!]}}
\newcommand{\etal}{{\it et al.}}
\newcommand{\cm}{cm$^{-1}$}
\newcommand{\pprime}{{\prime\prime}}
\newcommand{\ls}[4]{\ensuremath{^{#1}\!{#2}_{#3}^{#4}}}
\newcommand{\jj}[4]{\ensuremath{\left({\textstyle #1,#2}\right)_{#3}^{#4}}}
\newcommand{\s}{\ls{1}{S}{0}{}}
\newcommand{\p}{\ls{3}{P}{0}{}}
\newcommand{\pa}{\ls{3}{P}{1}{}}
\newcommand{\yb}[1]{$^{#1}$Yb}
\newcommand{\clocks}{\ensuremath{6s^2\,\ls{1}{S}{0}{}}}
\newcommand{\clockp}{\ensuremath{6s6p\,\ls{3}{P}{0}{}}}

\newcolumntype{.}{D{.}{.}{-1}}


\newcommand{\odd}[1]
{%
  \ifnum#1=1{\ensuremath{6s6p\,{\ls{3}{P}{1}{}}}}\fi%
  \ifnum#1=2{\ensuremath{6s6p\,{\ls{1}{P}{1}{}}}}\fi%
  \ifnum#1=3{\ensuremath{(4f^{13})5d6s^2\,\jj{\frac{7}{2}}{\frac{5}{2}}{1}{}}}\fi%
}

\newcommand{\even}[1]
{%
  \ifnum#1=1{\ensuremath{5d6s\,{\ls{3}{D}{1}{}}}}\fi%
  \ifnum#1=2{\ensuremath{6s7s\,{\ls{3}{S}{1}{}}}}\fi%
  \ifnum#1=3{\ensuremath{6s6d\,{\ls{3}{D}{1}{}}}}\fi%
  \ifnum#1=4{\ensuremath{6s8s\,{\ls{3}{S}{1}{}}}}\fi%
  \ifnum#1=5{\ensuremath{(4f^{13})5d6s6p\,{\jj{\frac{7}{2}}{\frac{5}{2}}{1}{}}}}\fi%
  \ifnum#1=6{\ensuremath{6p^2\,{\ls{3}{P}{1}{}}}}\fi%
  \ifnum#1=56{mixed states}\fi%
}


\newcommand{\NIST}{
National Institute of Standards and Technology, 
325 Broadway, Boulder, Colorado 80305, USA}

\newcommand{\CU}{
University of Colorado, Department of Physics,
Boulder, Colorado 80309, USA}

\newcommand{\UNR}{
University of Nevada, Department of Physics,
1664 N. Virginia St., Reno, Nevada 89512, USA }

\title{Determination of the $5d6s\,{}^3\!D_1$ state lifetime and blackbody radiation clock shift in Yb}

\author{K. Beloy}\email{kyle.beloy@nist.gov}
\affiliation{\NIST}
\author{J. A. Sherman}\email{jeff.sherman@nist.gov}
\affiliation{\NIST}
\affiliation{\CU}
\author{N. D. Lemke}
\affiliation{\NIST}
\affiliation{\CU}
\author{N. Hinkley}
\affiliation{\NIST}
\affiliation{\CU}
\author{C. W. Oates}
\affiliation{\NIST}
\author{A. D. Ludlow}
\affiliation{\NIST}

\date{\today}

\begin{abstract}
The Stark shift of the ytterbium optical clock transition due to room temperature blackbody radiation is dominated by a static Stark effect, which was recently measured to high accuracy [J.~A.~Sherman \emph{et al.}, Phys.\ Rev.\ Lett.\ \textbf{108}, 153002 (2012)].  However, room temperature operation of the clock at $10^{-18}$ inaccuracy requires a dynamic correction to this static approximation.  This dynamic correction largely depends on a single electric dipole matrix element for which theoretically and experimentally derived values disagree significantly.  We determine this important matrix element by two independent methods, which yield consistent values.  Along with precise radiative lifetimes of \odd{1} and \even{1}, we report the clock's blackbody radiation shift to 0.05\% precision.
\end{abstract}


\maketitle

%

Alkaline-earth-like atoms, such as Yb~\cite{LemLudBar09etal}, Sr~\cite{LeTBaiFou06etal,LudZelCam08etal,FalSchWin11etal}, and Hg~\cite{McFYiMej12etal} feature intrinsically narrow ${}^1\!S_0 \leftrightarrow {}^3\!P_0$ optical transitions capable of serving as stable and accurate frequency references~\cite{DerKat11} when cooled and held in an optical lattice trapping potential~\cite{PorDerFor04,TakHonHig05}. Accurate knowledge of clock transition frequencies  advances timekeeping technology and enables new tests of physics~\cite{RosHumSch08etal,BlaLudCam08etal,ChoHumRos10}.

Atomic frequency references are defined by an ideal system:  atoms at rest in a null-field, zero-temperature environment~\cite{TayTho08}. If a physical realization deviates from these ideals, researchers must account for corrections to the measured transition frequency and, importantly, uncertainty present in these corrections. Here, we explore the dominant ytterbium clock correction~\cite{LemLudBar09etal,SheLemHin12etal} due to room-temperature blackbody radiation (BBR).

\begin{figure}
\includegraphics[width=1.0\linewidth]{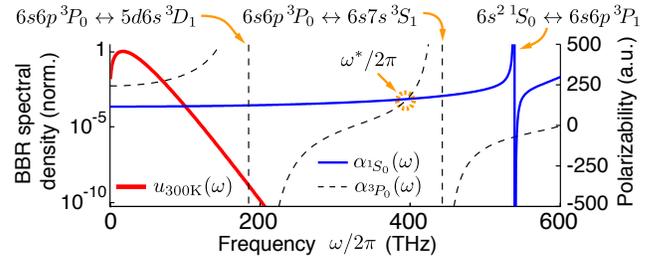}
\caption{(color online) Most of the room-temperature BBR energy spectrum (thick red line) is far infrared of transitions involving the Yb clock states.  The states' polarizabilities [see~Eq.~(\ref{eq:dynamicPolDef})], largely constant over much of the BBR spectrum, are balanced at the `magic' trapping frequency $\omega^*$.
}
\label{fig:bbrToyPlot}
\end{figure}

The polarizing effect of BBR largely mimics that of a static electric field due to the low frequency nature of BBR relative to optical transitions involving clock states (see Fig.~\ref{fig:bbrToyPlot}).  Writing the BBR clock frequency shift~\cite{PorDer06}
\begin{align} \label{eq:simpleBBR}
\Delta \nu_\text{BBR} &= - \frac{1}{2} \frac{\Delta \alpha(0)}{h} \langle E^2 \rangle_T [1 + \eta_\text{clock}(T)], \\
\intertext{highlights its similarity to a static Stark shift, where}
\label{eq:staticPol}
\Delta \alpha(0) &\equiv \alpha_e(0) - \alpha_g(0) = 145.726(3) \, \mathrm{a.u.}, 
\end{align}
is the differential \emph{static} polarizability between clock states $|g \rangle \equiv |6s^2\,{}^1\!S_0 \rangle$ and $|e \rangle \equiv |6s6p\,{}^3\!P_0 \rangle$, now known to high accuracy~\cite{SheLemHin12etal} (a.u.\ = atomic units~%
\footnote{The atomic unit for polarizability is $1 \text{ a.u.} = \text{0.248 8319} \times h\times\text{mHz}(\text{V/cm})^{-2}$, where $h$ is Planck's constant.  Expressed in atomic units, $e = m_e = \hbar = 4 \pi \epsilon_0 = 1$ \cite{MitSafCla10}.}%
). $\langle E^2 \rangle_T \approx (8.3193 \text{ V/cm})^2 (T/300 \text{ K})^4$ is the time-averaged electric field intensity of BBR at absolute temperature $T$~\cite{AngDzuFla06a}.  A small dynamic correction $\eta_\text{clock}(300~\text{K}) < 0.02$ accounts for the frequency dependence of $\Delta \alpha(\omega)$. 
Over 90\% of $\eta_\text{clock}$ depends on coupling between $6s6p\,{}^3\!P_0$ and neighboring $5d6s\,{}^3\!D_1$~(see~\cite{PorDer06} and supplemental material (SM)). But, critically, a measurement~\cite{BowBudCom96etal} and recent precise calculation~\cite{DzuDer10} of the electric dipole matrix element
\begin{equation*}
\mathcal{D} \equiv |\langle6s6p\,{}^{3}\!P_0 || \mathbf{D} ||5d6s\,{}^{3}\!D_1 \rangle|,
\end{equation*}
differ by $\sim \!3$ standard deviations. 

In this work, we independently determine $\mathcal{D}$ in order to accurately compute $\eta_\text{clock}$ and the clock correction $\Delta \nu_\text{BBR}$.  We present two distinct approaches resulting in good agreement.  First, we describe a semi-empirical technique which combines existing polarizability data with atomic theory to constrain $\mathcal{D}$.  Then we describe a measurement of the $5d6s\,{}^{3}\!D_1$ radiative lifetime in trapped Yb.  Since $5d6s\,{}^{3}\!D_1$ decays predominantly to the $6s6p\,{}^{3}\!P_J$ manifold, $\mathcal{D}$ is readily extracted. 
Finally, we discuss current accuracy limitations imposed by $\Delta \nu_\text{BBR}$.

\emph{Method I: Semi-empirical technique---}%
Accurately measured experimental parameters, such as the differential static polarizability [Eq.~(\ref{eq:staticPol})], also depend on electric dipole coupling between $6s6p\,{}^3\!P_0$ and $5d6s\,{}^3\!D_1$, and subsequently can be used to constrain the value of $\mathcal{D}$~\cite{PorLudBoy08}.
The electric dipole polarizability of clock state $n$ due to radiation at an off-resonant frequency $\omega/2\pi$ reads
\begin{equation}\label{eq:dynamicPolDef}
\alpha_n(\omega) = \frac{1}{\hbar}\frac{2}{3} \sum_{n' \ne n} \left| \langle n' || \mathbf{D} || n \rangle \right|^2 \frac{\omega_{n'n}}{\omega_{n'n}^2 - \omega^2},
\end{equation}
where $\langle n' || \mathbf{D} || n \rangle$ is a reduced electric dipole matrix element and $\omega_{n'n}/2\pi = \left(W_{n'} - W_{n}\right)/h$ is the corresponding transition frequency.  As $\omega \to 0$, we recover the familiar static polarizability expression.

The so-called `magic' trapping frequency $\omega^*$, which balances the polarizabilities of the clock states~\cite{DerKat11},
\begin{equation} \label{eq:dynamicPol}
\Delta \alpha(\omega^*) \equiv \alpha_e(\omega^*) - \alpha_g(\omega^*) = 0,
\end{equation}
has been measured to high accuracy in Yb \cite{LemLudBar09etal,BarStaLem08etal}.
Equations (\ref{eq:staticPol}) and (\ref{eq:dynamicPol}) may be combined to yield
\begin{eqnarray}
\Delta \alpha(0)+b\,\Delta \alpha(\omega^*)=145.726(3)~\mathrm{a.u.},
\label{Eq:Deltapsi}
\end{eqnarray}
where $b$ is arbitrary but may be chosen to 
to our benefit.
In the linear combination
\begin{eqnarray}
\alpha_n(0)+b\,\alpha_n(\omega^*)&=&
\frac{1}{\hbar}\frac{2}{3}\sum_{n' \ne n} \left| \langle n' || \mathbf{D} || n \rangle \right|^2\frac{1}{\omega_{n^\prime n}}
\nonumber\\&&\times
\left(
1+b\frac{\omega_{n^\prime n}^2}{\omega_{n^\prime n}^2-\omega^{*2}}
\right),
\label{Eq:weights}
\end{eqnarray}
the term in parenthesis serves as a `scale factor' relative to each transition's static polarizability contribution.  For instance, for a choice $b = -1$, this scale factor tends to zero for  $\omega_{n^\prime n}\gg\omega^*$ because these transitions contribute nearly identically to both polarizabilities $\alpha_n(0)$ and $\alpha_n(\omega^*)$ [see Eq.~(\ref{eq:dynamicPolDef})].
We find advantage in choosing a value $b \approx -1$ such that contributions from certain low-lying transitions are suppressed in the linear combination $\Delta \alpha(0) + b\, \Delta \alpha(\omega^*)$, along with contributions from the higher-lying transitions.
In Table \ref{Tab:alphas} we present contributions to Eq.~(\ref{Eq:weights}) from the lowest-lying transitions in each clock state for both $b=0$ and $b=-0.75$. In each case, we write the contribution from 
 $\clockp\rightarrow\even{1}$ in terms of unspecified matrix element $\mathcal{D}$. Other contributions are derived from experimental lifetimes in Refs.~\cite{TakKomHon04etal,BlaKom94,BauBraGai85}. Transitions to the closely spaced states $\even{5}$ and $\even{6}$ (`mixed states' in Table \ref{Tab:alphas}) are exceptions; these contributions were estimated with a CI+MBPT calculation similar to Ref.~\cite{DzuDer10}.
It is evident from Table \ref{Tab:alphas} how the choice of $b$ affects the relative importance of certain transitions. For example, while the `mixed states' contribute sizably to the differential static polarizability $\Delta\alpha(0)$, their contribution to $\Delta \alpha(0)-0.75\Delta \alpha(\omega^*)$ is negligible.
Moreover, contributions from higher-lying transitions not explicitly shown in the table---which contribute at the $\sim\!10\%$ level for both state polarizabilities \cite{pollimitms}---are also largely suppressed with the choice $b=-0.75$. Specifically, the `scale factor' in Eq.~(\ref{Eq:weights}) is nearly zero for the lowest of these transitions (for which $\omega_{n^\prime n}\approx 2\,\omega^*$), rising to just 0.25 for the highest-lying, least important transitions.

\begin{table}
\caption{Contributions to the static polarizability $\alpha_n(0)$ and linear combination $\alpha_n(0)-0.75\alpha_n(\omega^*)$ [refer to Eq.~(\ref{Eq:weights})] for the lowest-lying transitions from the clock states (a.u.). 
\label{Tab:alphas}}
\begin{ruledtabular}
\begin{tabular}{lcc}
$n^\prime$ 
& $\alpha_n(0)$ 
& $\alpha_n(0)-0.75\alpha_n(\omega^*)$ 
\\
\hline
\vspace{-2mm}\\
\multicolumn{1}{c}{}&
\multicolumn{2}{c}{clock state $n=\clocks$}\\
\odd{1}& 2 	& $-1$ \\
\odd{2}& 100 	& $-4$ \\
\odd{3}& 21 	& 1 \\
\vspace{-1mm}\\
\multicolumn{1}{c}{}&
\multicolumn{2}{c}{clock state $n=\clockp$}\\
\even{1}& $20.3\mathcal{D}^2$ 	& $26.8\mathcal{D}^2$ \\
\even{2}& 37 	& $-65$ \\
\even{3}& 22 	& $-3$ \\
\even{4}& 2 	& $0$ \\
\even{56}& 39 	& $0$ \\
\vspace{-4mm}
\end{tabular}
\end{ruledtabular}
\end{table}

Tallying contributions from all transitions, we find
\begin{eqnarray}
\Delta \alpha(0)-0.75\Delta \alpha(\omega^*)=26.8\mathcal{D}^2-64(8)+0(6),
\label{Eq:rhs}
\end{eqnarray}
in atomic units.
Here the first two terms on the r.h.s.~account for contributions from all transitions in Table \ref{Tab:alphas}; the uncertainty is dominated by that of the matrix element $\langle\even{2}||\mathbf{D}||\clockp\rangle$~\cite{BauBraGai85}.
The additional term 0(6) accounts for contributions from all higher-lying transitions not given explicitly in Table~\ref{Tab:alphas}. We ascribe an uncertainty to this term based on experimental upper limits to the polarizabilities of the two clock states \cite{pollimitms}, along with theoretical input from Ref.~\cite{DzuDer10} and present CI+MBPT calculations. Equating the r.h.s.~of (\ref{Eq:rhs}) to experimental result (\ref{Eq:Deltapsi}) gives 
$
\mathcal{D}=2.80(7)~\mathrm{a.u.} 
$
We compare this result with other determinations and new data below.

\begin{figure}[t]
\includegraphics[width=\linewidth]{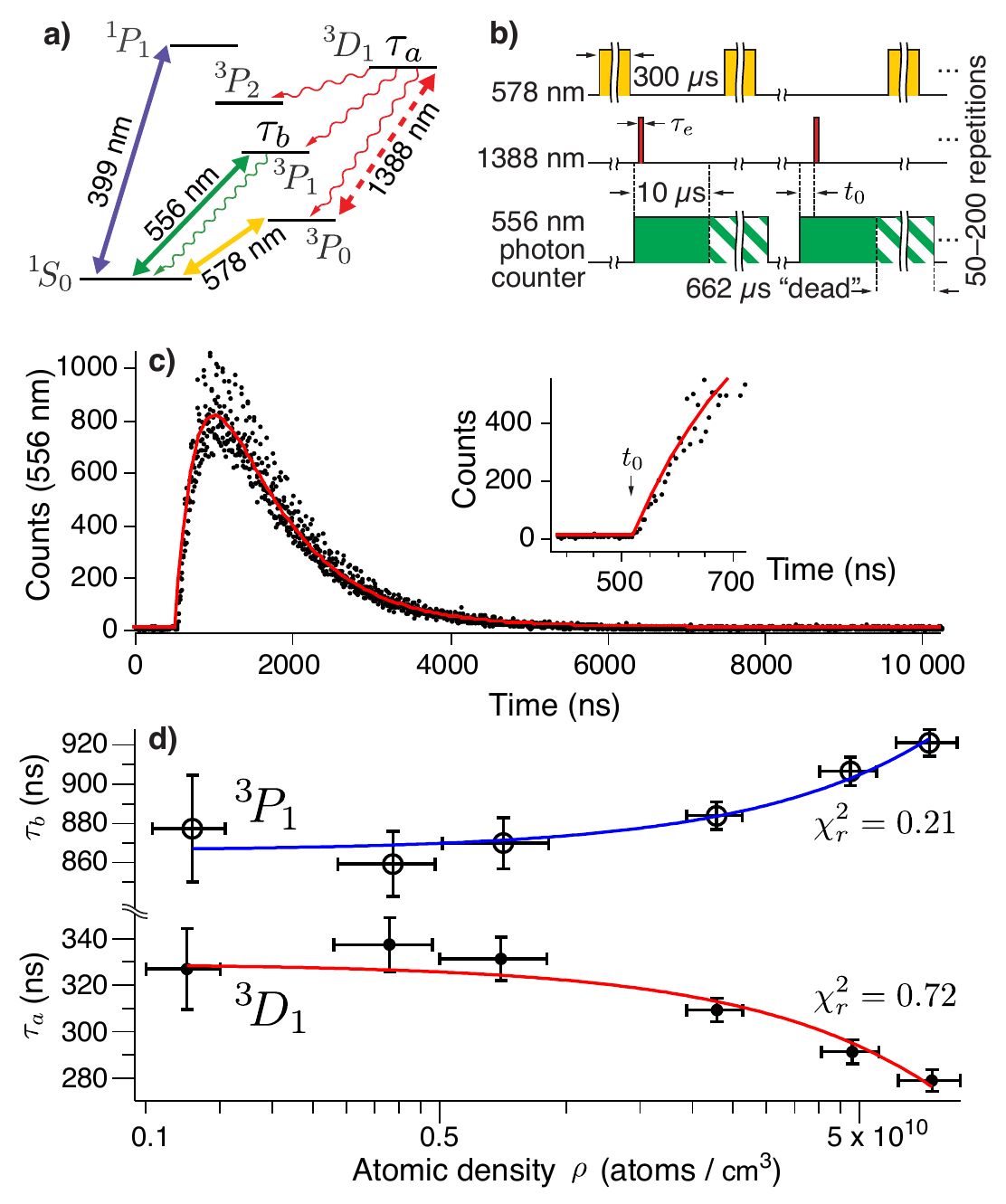}
\caption{(color online) We measure the $5d6s\,{}^3\!D_1$ radiative lifetime via cascade decay. a) Key lifetimes are labeled $\tau_a$ and $\tau_b$.  Double-arrowed lines indicate laser transitions.  Wavy lines are important decay channels.  b) A representative pulse timing diagram.  c) Typical fluorescence data, fit (red line) to Eq.~(\ref{eq:decayEq}).  An inset highlights the signal and fit near an excitation time $t_0 = 523.3$(6)~ns.  d) Observed lifetimes vary with atomic density, $\rho$.  Lines are linear regression fits.  Logarithmic scaling emphasizes data at low $\rho$ with negligible interaction effects. Error bars represent standard uncertainties obtained from nonlinear fits, and uncertainty in $\rho$ estimates.}
\label{fig:experimentalStuff}
\end{figure}

\emph{Method II: lifetime measurement---}%
Alternatively, measurement of the $5d6s\,{}^{3}\!D_1$ radiative lifetime $\tau_a$ yields $\mathcal{D}$ since $\mathcal{D}^2 =  (3 \pi \epsilon_0 \hbar c^3 \zeta_0)(2J' +1) / (\omega_0^3 \tau_a)$, where $J'=1$; $\omega_0 / 2\pi \approx 2.1587 \times 10^{14}$~Hz and $\zeta_0 = 0.64(1)$ are the radiated frequency and branching ratio to ${}^3\!P_0$, respectively.  $\zeta_0$ is accurately computable because $LS$-coupling remains valid~\cite{MarZalHag78}.  In the cascade $5d6s\,{}^3\!D_1 \to 6s6p\,{}^3\!P_1 \to 6s^2\,{}^1\!S_0$ [see Fig.~\ref{fig:experimentalStuff}(a)], atoms emit a 556~nm photon during the second decay which is technically easier to detect than the first radiated (infrared) photon~\cite{BowBudCom96etal}.  Other states populated by the decay ($^{3}\!P_0$ and $^{3}\!P_2$) are long-lived.  If atoms are instantaneously excited at time $t_0$ to $^{3}\!D_1$, fluorescence from $^{3}\!P_1$ follows a double exponential~\cite{BudDeMCom94, BowBudCom96etal},
\begin{equation}
y(t) = A \times \Theta(t - t_0) \left[e^{-(t-t_0)/\tau_a} - e^{-(t-t_0)/\tau_b} \right] + y_0,
\label{eq:decayEq}
\end{equation}
where $\tau_b$ is the radiative lifetime of ${}^{3}\!P_1$ ($\tau_b > \tau_a$),  $A$ is a scaling factor and $y_0$ accounts for stray detected light. The Heaviside unit-step $\Theta(t- t_0)$ models rapid atom excitation at $t_0$.   Decay branching ratios affect only the normalization of Eq.~(\ref{eq:decayEq}), \emph{not} its time dependence~\cite{BudKimDeM}.

We describe the cooling and confinement of $N_\text{at} \approx 10^4$ atoms of  ${}^{171}$Yb in a one-dimensional optical lattice elsewhere~\cite{LemLudBar09etal}.  As depicted in Fig.~\ref{fig:experimentalStuff}(b), a resonant `$\pi$-pulse' of 578~nm light~\cite{JiaLudLem11etal} coherently transfers atoms from $^{1}\!S_0$ to the long-lived $^{3}\!P_0$ state.  Then, a brief ($\tau_e = 25$~ns) resonant pulse of 1388~nm light excites more than half of these atoms to $^{3}\!D_1$.  An event counter accumulates the arrival times of radiated 556~nm photons into 5~ns bins.  Heating due to photon scattering, background gas collisions, and accumulation in $^{3}\!P_2$ limit the number of excitations per loading cycle to about 200.  Though we estimate a modest light collection/detection efficiency ($\sim 0.1\,\%$), we typically observe $N_\text{at} \times (1.5 \times 10^{-5})$ green photons per excitation.  Biases due to photon `pile-up' in counter bins are negligible.  Between $10^6$--$10^7$ excitations are sufficient to obtain satisfactory decay profiles [e.g., Fig.~\ref{fig:experimentalStuff}(c)].

\begin{figure}
\centering
\includegraphics[width=1.0\linewidth]{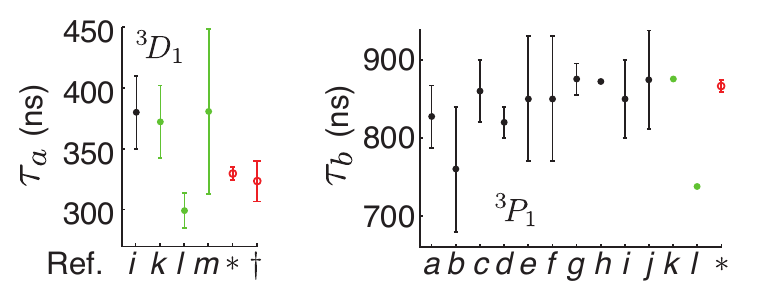}
\caption{(color online) The present measurement ($*$) and semi-empirical result ($\dag$) (open red circles) are compared to other determinations.  Measurements \emph{a--j} are from, respectively, \cite{BauWan66,BudSni70,GorKaiLan72,RamSch76,BurVerKom74,BlaKomPen78,GusLunNil79,GolBaiMos88}, \cite{BowBudCom96etal}, \cite{KitEnoKas08}.  We note that the small error bar on \emph{h} accounts only for statistical uncertainty.  Green points \emph{k--m} are calculations~\cite{PorRakKoz99}, \cite{DzuDer10}, \cite{GuoWanYe10}.  When necessary, we infer lifetimes from reported matrix elements or natural linewidths.  We assign an error bar to the $\tau_a$ prediction of \emph{l} using the authors' estimate of uncertainty in a highly correlated polarizability.}
\label{fig:results}
\end{figure}

We fit fluorescence signals to Eq.~(\ref{eq:decayEq}) with a statistically-weighted Levenberg-Marquardt routine. Though covariance in $A$, $\tau_a$, and $\tau_b$ can be significant, simulations establish that fitting biases become negligible with sufficient count totals.  In large data sets, event counter technical noise synchronous with a timing oscillator overwhelms the signal shot noise.  Re-binning data into 20~ns chunks removes much of this noise, but does not significantly alter the results.  A maximum likelihood method yielded statistically similar fits.

Atomic interactions, such as collective emission (`superradiance', `subradiance') or radiation trapping may influence radiative decay.  We probed these effects by varying the atomic density $\rho$ undergoing decay~\footnote{Here, $\rho$ is defined as the mean density during the repeated cascade decay sequence~[see Fig.~\ref{fig:experimentalStuff}(b)] and is measured by alternately inserting pulses of resonant 399~nm light before and after the sequence.  We assume fluorescence, properly normalized against background light, is $\propto \rho$.}.  Results [Fig.~\ref{fig:experimentalStuff}(d)] indicate non-negligible shortening of $\tau_a$ at high $\rho$. In the limit of slow dipole dephasing~\footnote{The dephasing timescale in cold trapped atoms is likely set by inhomogeneous ac-Stark shifts. We estimate $T_2^*\approx 5~\mu$s, which is somewhat longer than $\tau_a$ and $\tau_b$.}, collective emission shortens an observed decay lifetime as $\tau = \tau_0 (1+ \rho L \lambda^2/4)^{-1}$,
where $\tau_0$ is the single atom value, $L$ is the length of a pencil-shaped atomic cloud, and $\lambda$ is the radiated wavelength~\cite{GroFabPil76}.  For $\rho = 10^{9} \text{ cm}^{-3}$ and $L = 0.1$~mm, the modification in this simple model is about 5\%.   We see the opposite effect in ${}^3\!P_1$, an \emph{increase} of $\tau_b$ at high $\rho$.  We explored both effects by varying 578~nm and 1388~nm excitation pulse areas, altering the relative populations of ${}^1\!S_0$, ${}^3\!P_0$, and ${}^3\!D_1$, but observed no substantial change in the systematic effect;  further investigation is warranted.  Nevertheless, the effects of interactions on observed lifetimes are made negligible over the lowest decade of examined densities [see Fig.~\ref{fig:experimentalStuff}(d)].  Quenching due to cold-collisions and lattice scattering is negligible. 

\begin{table}[t!]
\caption{The uncertainty in lifetimes $\tau_a$ and $\tau_b$ due to atomic interactions is largely statistical since we report extrapolations to zero atomic density.  Covariant fit biases and distortions due to Zeeman oscillations are estimated from Monte-Carlo simulations.  Uncertainties due to 1388~nm pulse duration ($\tau_e$) and stray light are statistically limited.}
\begin{ruledtabular}
\begin{tabular}{p{1.6 in}cc}
								& $u(\tau_a)$ (ns) 	& $u(\tau_b)$ (ns) \\ \hline
Atomic interactions					& 4.3				&  3.3	\\
Fit biases							& 0.9				& 1.5		\\
Zeeman quantum beats				& 3.0				& 3.0		\\
1388~nm finite excitation				& 3.3				& 4.3		\\
1388~nm stray  light	 				& 3.4				& 4.6		\\
Event counter timing 				& 0.2 			&  0.4	\\ \hline
Total (quadrature sum)				& 7.1				& 7.4
\end{tabular}
\end{ruledtabular}
\label{tab:systematics}
\end{table}



Potential systematic effects arise from finite 1388~nm pulse duration $\tau_e$, and spurious excitation due to poor extinction.  We varied $\tau_e$ from 25~ns to 90~ns and observed no significant systematic effect on fitted state lifetimes.  We configured the apparatus for fast actuation and high ($60\,$dB) extinction of the 1388~nm light.  A single-mode fiber-coupled mechanical shutter ($500\,\mu$s rise-time) precedes a fiber-coupled acousto-optic modulator (AOM).  A second (free-space) AOM, driven with a tuned delay, aids in pulse shaping and extinction.  A beam dump, iris, and baffles minimize the influence of scattered light.  The final AOM imposes a 160~MHz frequency shift, detuning scattered light from resonance and increasing effective extinction. With up to 1~mW of deflected 1388~nm light focused to a 30~$\mu$m spot (an intensity $I = 7 \times 10^5$~W/m$^2$), we attain a Rabi frequency $\Omega \propto \sqrt{I} \mathcal{D}$ exceeding 1~GHz.  We observed no significant changes to $\tau_a$ and $\tau_b$ when varying an optical attenuator over 4~dB to test for dependence on $\Omega$ and stray light intensity.

We systematically excited to both hyperfine components ${}^3\!D_1(F' = \tfrac{1}{2}, \tfrac{3}{2})$, which are split by 3.07(7) GHz. We observe no hyperfine quantum beats~\cite{HarPaiSch73} due to the large splitting and selective laser excitation.  We observed no Zeeman oscillations~\cite{BowBudCom96etal} or significantly different results when an applied magnetic field $\vec{B}$ varied from 0.01~mT to 0.1~mT.  1388~nm excitation light propagated along the lattice axis and was polarized perpendicular to $\vec{B}$.  We detected 556~nm photons $\approx 45^\circ$ from the lattice axis with largely polarization insensitive optics.  We observed a slight dependence on the number of excitations per loading cycle but believe this effect is due to atomic interactions since increased scattering reduces the atom number.  Varying the lattice laser intensity over 50\% yielded no significant change in observed lifetimes.

We report the radiative lifetimes
$\tau_a  = (329.3 \pm 7.1) \text{ ns}$ and
$\tau_b  = (866.1 \pm 7.4) \text{ ns}$.
Table~\ref{tab:systematics} enumerates measurement uncertainties. We compare our results to existing measurements and calculations in Fig.~\ref{fig:results}.  Our result for $\tau_b$ agrees with many prior measurements.  Our result for $\tau_a$ lies between the only other measurement~\cite{BowBudCom96etal} and a recent calculation~\cite{DzuDer10}.  Our semi-empirical method exhibits good agreement with the measurement. Table~\ref{tab:matrixElementResults} lists the results as inferred matrix elements.


\begin{table}
\caption{The present results, expressed as reduced matrix elements (a.u.), are compared with selected literature values.}
\begin{ruledtabular}
\begin{tabular}{p{0.70in}..}
			& \multicolumn{1}{c}{$\mathcal{D}$}
	 		& \multicolumn{1}{c}{$\left| \langle6s6p\,{}^{3}\!P_1 || \mathbf{D} ||6s^2\,{}^{1}\!S_0 \rangle \right|$} \\ \hline
\raggedright \mbox{Experiment} 			& 2.77(4)  				& 0.542(2) \\
\raggedright \mbox{Semi-empirical} 			& 2.80(7) 				& \multicolumn{1}{c}{---} \\ \hline
Experiment~\cite{BowBudCom96etal}		& 2.58(10)			& 0.547(16) 	\\
Calculation~\cite{PorRakKoz99}			& 2.61(10)			& 0.54(8)		\\
Calculation~\cite{DzuDer10}				& 2.91(7)				& 0.587		\\
Calculation~\cite{GuoWanYe10}			& 2.58(23)			& 0.41(1)		
\end{tabular}
\end{ruledtabular}
\label{tab:matrixElementResults}
\end{table} 

{\it BBR clock shift---}%
The BBR Stark shift to the clock frequency is found from the expression
\begin{eqnarray}
\Delta \nu_\text{BBR}=
-\frac{1}{2h\epsilon_0}\,
\int_0^\infty u_T(\omega)\,\Delta\alpha(\omega)\,d\omega,
\label{Eq:BBRshiftfull}
\end{eqnarray}
where $u_T(\omega)$ is the BBR spectral energy density corresponding to temperature $T$, given by Planck's law. 
A static approximation neglecting the slight frequency dependence of $\Delta \alpha(\omega)$ over the BBR spectrum (refer to Fig.~\ref{fig:bbrToyPlot}) is formally obtained by making the substitution $\Delta\alpha(\omega)\rightarrow\Delta\alpha(0)$ in Eq.~(\ref{Eq:BBRshiftfull}).
An improved approximation takes into account the lowest-order frequency dependence of the polarizability arising from the low-lying $\clockp\rightarrow\even{1}$ transition: 
$\Delta\alpha(\omega)\rightarrow\Delta\alpha(0)+(2/3\hbar)(\mathcal{D}^2/\omega_0^3)\omega^2$. Integrating over $\omega$ analytically, we interpret the additional shift as 
$
\eta_\mathrm{clock}(T)\approx\frac{80\pi^2}{63}\frac{\mathcal{D}^2}{(\hbar\omega_0)^3}\frac{(k_BT)^2}{\Delta\alpha(0)}\approx 0.017\left(\frac{T}{300~\mathrm{K}}\right)^2
$
from Eq.~(\ref{eq:simpleBBR}), where $k_B$ is Boltzmann's constant.
A more thorough account of small contributions from all other transitions, including the ${}^1\!S_0$ state and next-order terms ($\propto T^4$) yields
\begin{eqnarray*}
\eta_\mathrm{clock}(T)=
0.0173(5)\left(\frac{T}{300~\mathrm{K}}\right)^2
+0.0006\left(\frac{T}{300~\mathrm{K}}\right)^4.
\end{eqnarray*}
We omit higher order terms ($\propto T^6,T^8,\dots$) which are negligible at $T \lesssim 300$~K. 
We provide more details of this evaluation, including multipolar effects~\cite{PorDer06}, in the SM.



{\it Conclusion---}%
Assuming an ideal BBR environment at 300~K, we use the present results to calculate $\Delta \nu_\text{BBR} = -1.2774(6)$~Hz. The present determination of $\eta_\text{clock}$ sets an uncertainty limit for $\Delta \nu_\text{BBR}$ at $1.1 \times10^{-18}$.  In practice, uncertainty in $\Delta \nu_\text{BBR}$ also arises from imprecise knowledge of the thermal environment. In an existing apparatus, we estimate an effective temperature uncertainty of 1 K due to non-uniformity, corresponding to a fractional clock uncertainty of $3.3 \times 10^{-17}$~\cite{LemLudBar09etal}. The present results therefore motivate further efforts to control the thermal environment of the clock chamber~\cite{MidLisFal11}.

{\it Acknowledgements---}The authors acknowledge NIST, DARPA QuASar, NASA, and the NRC RAP program for financial support. We thank J.\ Ye and J.\ Bergquist for equipment loans; A.\ Derevianko and V.\ Dzuba provided useful discussion.  We thank N.\ Phillips for laboratory contributions.

%

\newpage
\begin{widetext}

\newcommand{\pick}[3]{%
 \ifnum#2=\level%
  \ifnum#1=\parity%
   {#3}%
  \fi%
 \fi}
\newcommand{\parity}{0}
\newcommand{\level}{0}


 \newcommand{\state}[2]{%
 \renewcommand{\parity}{#1}%
 \renewcommand{\level}{#2}%
 \pick{1}{1}{\ensuremath{6s6p\,{\ls{3}{P}{1}{}}}}%
 \pick{1}{2}{\ensuremath{6s6p\,{\ls{1}{P}{1}{}}}}%
 \pick{1}{3}{\ensuremath{(4f^{13})5d6s^2\,\jj{\frac{7}{2}}{\frac{5}{2}}{1}{}}}%
 \pick{1}{99}{all others}%
 \pick{0}{1}{\ensuremath{5d6s\,{\ls{3}{D}{1}{}}}}%
 \pick{0}{2}{\ensuremath{6s7s\,{\ls{3}{S}{1}{}}}}%
 \pick{0}{3}{\ensuremath{6s6d\,{\ls{3}{D}{1}{}}}}%
 \pick{0}{4}{\ensuremath{6s8s\,{\ls{3}{S}{1}{}}}}%
 \pick{0}{5}{\ensuremath{(4f^{13})5d6s6p\,{\jj{\frac{7}{2}}{\frac{5}{2}}{1}{}}}}%
 \pick{0}{6}{\ensuremath{6p^2\,{\ls{3}{P}{1}{}}}}%
 \pick{0}{56}{mixed states}%
 \pick{0}{99}{all others}%
 \pick{9}{0}{$\Delta$(main)}%
 \pick{9}{56}{$\Delta$(mixed states)}%
 \pick{9}{99}{$\Delta$(all others)}%
 \pick{9}{-1}{$\Delta$(total)}%
 \pick{9}{-8}{$\Delta$(expt.)}%
}

\newcommand{\stated}{\ensuremath{5d6s\,{\ls{3}{D}{1}{}}}}

\newcommand{\freq}[2]{%
 \renewcommand{\parity}{#1}%
 \renewcommand{\level}{#2}%
 \pick{1}{1}{0.08198}%
 \pick{1}{2}{0.11422}%
 \pick{1}{3}{0.13148}%
 \pick{0}{1}{0.03281}%
 \pick{0}{2}{0.07020}%
 \pick{0}{3}{0.10261}%
 \pick{0}{4}{0.11084}%
 \pick{0}{5}{0.11458}%
 \pick{0}{6}{0.12082}%
}

\newcommand{\dip}[2]{%
 \renewcommand{\parity}{#1}%
 \renewcommand{\level}{#2}%
 \pick{1}{1}{$0.542\pm0.002$}%
 \pick{1}{2}{$4.148\pm0.002$}%
 \pick{1}{3}{$2.03\pm0.04$}%
 \pick{0}{1}{$2.77\pm0.04$}%
 \pick{0}{2}{$2.0\pm0.1$}%
 \pick{0}{3}{$1.82\pm0.03$}%
 \pick{0}{4}{$0.64\pm0.04$}%
}

\newcommand{\alphstat}[2]{%
 \renewcommand{\parity}{#1}%
 \renewcommand{\level}{#2}%
 \pick{1}{1}{$2.39\pm0.02$}%
 \pick{1}{2}{$100.40\pm0.09$}%
 \pick{1}{3}{$20.8\pm0.8$}%
 \pick{1}{99}{$\chi_1$}%
 \pick{0}{1}{$156\pm4$}%
 \pick{0}{2}{$37\pm4$}%
 \pick{0}{3}{$21.6\pm0.7$}%
 \pick{0}{4}{$2.4\pm0.3$}%
 \pick{0}{5}{}%
 \pick{0}{6}{}%
 \pick{0}{56}{$\chi_2$}%
 \pick{0}{99}{$\chi_3$}%
 \pick{9}{0}{$93\pm6$}%
 \pick{9}{56}{$39\pm13$}%
 \pick{9}{99}{$16\pm16$}%
 \pick{9}{-1}{$148\pm21$}%
 \pick{9}{-8}{$145.726\pm0.003$}%
}

\newcommand{\alphmag}[2]{%
 \renewcommand{\parity}{#1}%
 \renewcommand{\level}{#2}%
 \pick{1}{1}{$5.14\pm0.04$}%
 \pick{1}{2}{$138.7\pm0.1$}%
 \pick{1}{3}{$26\pm1$}%
 \pick{1}{99}{$(1.00,1.14)\chi_1$}%
 \pick{0}{1}{$-67\pm2$}%
 \pick{0}{2}{$136\pm16$}%
 \pick{0}{3}{$33\pm1$}%
 \pick{0}{4}{$3.4\pm0.4$}%
 \pick{0}{5}{}%
 \pick{0}{6}{}%
 \pick{0}{56}{$(1.33,1.38)\chi_2$}%
 \pick{0}{99}{$(1.00,1.31)\chi_3$}%
 \pick{9}{0}{$-64\pm16$}%
 \pick{9}{56}{$52\pm17$}%
 \pick{9}{99}{$20\pm20$}%
 \pick{9}{-1}{$8\pm31$}%
 \pick{9}{-8}{$0$}%
}

\newcommand{\psicomb}[2]{%
 \renewcommand{\parity}{#1}%
 \renewcommand{\level}{#2}%
 \pick{1}{1}{$-1.47\pm0.01$}%
 \pick{1}{2}{$-3.601\pm0.003$}%
 \pick{1}{3}{$1.10\pm0.04$}%
 \pick{1}{99}{$(0.15,0.25)\chi_1$}%
 \pick{1}{990}{2(2)}%
 \pick{0}{1}{$206\pm5$}%
 \pick{0}{2}{$-65\pm8$}%
 \pick{0}{3}{$-3.02\pm0.09$}%
 \pick{0}{4}{$-0.15\pm0.02$}%
 \pick{0}{5}{}%
 \pick{0}{6}{}%
 \pick{0}{56}{$(-0.033,0.004)\chi_2$}%
 \pick{0}{99}{$(0.02,0.25)\chi_3$}%
 \pick{9}{0}{$142\pm10$}%
 \pick{9}{56}{$0.0\pm0.5$}%
 \pick{9}{99}{$0\pm6$}%
 \pick{9}{-1}{$142\pm11$}%
 \pick{9}{-8}{$145.726\pm0.003$}%
}

\newcommand{\alphtwo}[2]{%
 \renewcommand{\parity}{#1}%
 \renewcommand{\level}{#2}%
 \pick{1}{1}{$0.355\pm0.003$}%
 \pick{1}{2}{$7.696\pm0.007$}%
 \pick{1}{3}{$1.20\pm0.05$}%
 \pick{1}{99}{$(0.00,0.03)\chi_1$}%
 \pick{0}{1}{$145\pm4$}%
 \pick{0}{2}{$7.4\pm0.9$}%
 \pick{0}{3}{$2.05\pm0.06$}%
 \pick{0}{4}{$0.20\pm0.02$}%
 \pick{0}{5}{}%
 \pick{0}{6}{}%
 \pick{0}{56}{$(0.07,0.08)\chi_2$}%
 \pick{0}{99}{$(0.00,0.07)\chi_3$}%
 \pick{9}{0}{$146\pm4$}%
 \pick{9}{56}{$2.7\pm0.9$}%
 \pick{9}{99}{$1\pm1$}%
 \pick{9}{-1}{$149\pm4$}%
}

\newcommand{\rhocomb}[2]{%
 \renewcommand{\parity}{#1}%
 \renewcommand{\level}{#2}%
 \pick{1}{1}{$-0.219\pm0.002$}%
 \pick{1}{2}{$-0.2760\pm0.0003$}%
 \pick{1}{3}{$0.063\pm0.002$}%
 \pick{1}{99}{$(0.000,0.005)\chi_1$}%
 \pick{1}{990}{}%
 \pick{0}{1}{$192\pm5$}%
 \pick{0}{2}{$-13\pm2$}%
 \pick{0}{3}{$-0.287\pm0.009$}%
 \pick{0}{4}{$-0.012\pm0.001$}%
 \pick{0}{5}{}%
 \pick{0}{6}{}%
 \pick{0}{56}{$(-0.0025,0.0003)\chi_2$}%
 \pick{0}{99}{$(0.000,0.005)\chi_3$}%
 \pick{9}{0}{$179\pm5$}%
 \pick{9}{56}{$0.1\pm0.1$}%
 \pick{9}{99}{$0.08\pm0.08$}%
 \pick{9}{-1}{$179\pm5$}%
}



\newcommand{\fnm}[1]{%
 \ifnum#1=0%
 \else%
  \footnotemark[#1]%
 \fi%
}

\newcommand{\tabline}[3]{
\state{#1}{#2} & 
\freq{#1}{#2} & 
\dip{#1}{#2}\fnm{#3} &
\alphstat{#1}{#2} &
\alphmag{#1}{#2} &
\psicomb{#1}{#2} &
\alphtwo{#1}{#2} &
\rhocomb{#1}{#2} \\
}

\begin{center}%
\textbf{{\large%
Supplemental Material
}}%
\end{center}%

\section{Tabulated results for atomic factors}

Table \ref{Tab:alphas} presents transition frequencies and dipole matrix elements for transitions from the clock states, as well as corresponding contributions to static polarizability $\alpha_n(0)$, polarizability at the magic frequency $\alpha_n(\omega^*)$, and the linear combination $\alpha_n(0)-0.75\alpha_n(\omega^*)$ for both clock states. Also provided are contributions to the atomic properties $\alpha_n^{(2)}$ and $\rho_n$, which are discussed in the following section.

In the upper portion of Table \ref{Tab:alphas}, explicit values are given where experimental data is available, with uncertainties being derived from the corresponding references. For remaining transitions, contributions to the static polarizability are given as unknown parameters $\chi_{1,2,3}$. Corresponding contributions to the other atomic properties are a written as a range $(a,b)\chi_{1,2,3}$, where the range $(a,b)$ is determined purely by the atomic spectrum and magic frequency. This helps illustrate the relative importance of the unknown contributions in each case. For example, while $\chi_2$ denotes the combined contribution to static polarizablity $\alpha_e(0)$ due to the `mixed states', the associated contribution to $\alpha_e(0)-0.75\alpha_e(\omega^*)$ is comparatively suppressed, necessarily being between $-0.033\chi_2$ and $0.004\chi_2$.

In the middle portion of Table \ref{Tab:alphas}, contributions to the differential properties [e.g., $\Delta\alpha(0)\equiv\alpha_e(0)-\alpha_g(0)$] are tallied. Here $\Delta$(`main states') incorporates all values given explicitly in the upper portion of the table. $\Delta$(`mixed states') is estimated by CI+MBPT calculations similar to those described in Ref.~\cite{DzuDer10}, whereas $\Delta$(`all others') is estimated with CI+MBPT calculations together with additional theoretical input from Refs.~\cite{DzuDer10,pollimitms}. Theoretical uncertainty for $\Delta$(`mixed states') and $\Delta$(`all others') is difficult to assess; the numbers given in the table represent reasonable estimates of this uncertainty. We reiterate that methods developed in the main text minimize the influence of these contributions.

In Table \ref{Tab:alphas}, contributions from the $\clockp\rightarrow\state{0}{1}$ transition are derived from our most accurate determination (see main text) of the matrix element $\mathcal{D}\equiv\left|\langle\clockp||\mathbf{D}||\state{0}{1}\rangle\right|=2.77(4)$. In Table \ref{Tab:PD}, we compare these results with contributions derived from other determinations of this matrix element.

\begin{table*}[b]
\caption{Transition frequencies, matrix elements, and contributions to various atomic properties of interest for the Yb clock states. The magic frequency is $\omega^*=0.06000$ \cite{LemLudBar09etal,BarStaLem08etal}. $(a,b)\chi$ denotes a range $(a\chi,b\chi)$. All values are in atomic units.
\label{Tab:alphas}}
\begin{ruledtabular}
\begin{tabular}{lccccccc}
$n^\prime$ 
& $\omega_{n^\prime n}$ 
& $\left|\langle n^\prime||\mathbf{D}||n\rangle\right|$ 
& $\alpha_n(0)$ 
& $\alpha_n(\omega^*)$ 
& $\alpha_n(0)-0.75\alpha_n(\omega^*)$ 
& $\alpha_n^{(2)}/10^3$ 
& $\rho_n/10^3$ 
\\
\hline
\vspace{-2mm}\\
\multicolumn{1}{c}{}&
\multicolumn{7}{c}{clock state $n=\clocks$}\\
\tabline{1}{1}{1}
\tabline{1}{2}{2}
\tabline{1}{3}{3}
\tabline{1}{99}{0}
\vspace{-1mm}\\
\multicolumn{1}{c}{}&
\multicolumn{7}{c}{clock state $n=\clockp$}\\
\tabline{0}{1}{1}
\tabline{0}{2}{4}
\tabline{0}{3}{5}
\tabline{0}{4}{4}
\tabline{0}{56}{0}
\tabline{0}{99}{0}
\hline
\vspace{-1.5mm}\\
\tabline{9}{0}{0}
\tabline{9}{56}{0}
\tabline{9}{99}{0}
\hline
\vspace{-1.5mm}\\
\tabline{9}{-1}{0}
\tabline{9}{-8}{0}
\end{tabular}
\footnotetext[1]{expt., present}
\footnotetext[2]{Ref.~\cite{TakKomHon04etal}}
\footnotetext[3]{weighted mean from four values compiled in Ref.~\cite{BlaKom94}}
\footnotetext[4]{Ref.~\cite{BauBraGai85}}
\footnotetext[5]{Ref.~\cite{BauGeiLie81}}
\end{ruledtabular}
\end{table*}

\newcommand{\PDref}[1]{%
\ifnum#1=1{unspecified}\fi%
\ifnum#1=2{expt., present}\fi%
\ifnum#1=3{semi-emp., present}\fi%
\ifnum#1=4{expt., Ref.~\cite{BowBudCom96etal}}\fi%
\ifnum#1=5{theor., Ref.~\cite{DzuDer10}}\fi%
\ifnum#1=6{theor., Ref.~\cite{GuoWanYe10}}\fi%
}

\newcommand{\PDdip}[1]{%
\ifnum#1=1{$\mathcal{D}$}\fi%
\ifnum#1=2{\dip{0}{1}}\fi%
\ifnum#1=3{$2.80\pm0.07$}\fi%
\ifnum#1=4{$2.58\pm0.10$}\fi%
\ifnum#1=5{$2.91\pm0.07$}\fi%
\ifnum#1=6{$2.58\pm0.23$}\fi%
}

\newcommand{\PDalphstat}[1]{%
\ifnum#1=1{$20.32\mathcal{D}^2$}\fi%
\ifnum#1=2{\alphstat{0}{1}}\fi%
\ifnum#1=3{$159\pm 8$}\fi%
\ifnum#1=4{$135\pm 10$}\fi%
\ifnum#1=5{$172\pm 8$}\fi%
\ifnum#1=6{$135\pm24$}\fi%
}

\newcommand{\PDalphmag}[1]{%
\ifnum#1=1{$-8.666\mathcal{D}^2$}\fi%
\ifnum#1=2{\alphmag{0}{1}}\fi%
\ifnum#1=3{$-68\pm 3$}\fi%
\ifnum#1=4{$-58\pm 4$}\fi%
\ifnum#1=5{$-73\pm 4$}\fi%
\ifnum#1=6{$-58\pm10$}\fi%
}

\newcommand{\PDpsicomb}[1]{%
\ifnum#1=1{$26.82\mathcal{D}^2$}\fi%
\ifnum#1=2{\psicomb{0}{1}}\fi%
\ifnum#1=3{$210\pm 11$}\fi%
\ifnum#1=4{$179\pm 14$}\fi%
\ifnum#1=5{$227\pm 11$}\fi%
\ifnum#1=6{$179\pm 32$}\fi%
}

\newcommand{\PDalphtwo}[1]{%
\ifnum#1=1{$18.88\mathcal{D}^2$}\fi%
\ifnum#1=2{\alphtwo{0}{1}}\fi%
\ifnum#1=3{$148\pm 7$}\fi%
\ifnum#1=4{$126\pm 10$}\fi%
\ifnum#1=5{$160\pm 8$}\fi%
\ifnum#1=6{$126\pm 22$}\fi%
}

\newcommand{\PDrhocomb}[1]{%
\ifnum#1=1{$24.92\mathcal{D}^2$}\fi%
\ifnum#1=2{\rhocomb{0}{1}}\fi%
\ifnum#1=3{$195\pm 10$}\fi%
\ifnum#1=4{$166\pm 13$}\fi%
\ifnum#1=5{$211\pm 10$}\fi%
\ifnum#1=6{$166\pm 30$}\fi%
}

\newcommand{\PDtabline}[1]{
\PDref{#1} & 
\PDdip{#1} &
\PDalphstat{#1} &
\PDalphmag{#1} &
\PDpsicomb{#1} &
\PDalphtwo{#1} &
\PDrhocomb{#1} \\
}

\begin{table*}
\caption{Comparison of the $\clockp\rightarrow\stated$ contribution to the atomic properties in Table \ref{Tab:alphas} with different determinations of the dipole matrix element. Here $e$ specifies the excited clock state \clockp. All values are in atomic units.
\label{Tab:PD}}
\begin{ruledtabular}
\begin{tabular}{lcccccc}
& $\left|\langle \clockp||\mathbf{D}||\stated\rangle\right|$ 
& $\alpha_e(0)$ 
& $\alpha_e(\omega^*)$ 
& $\alpha_e(0)-0.75\alpha_e(\omega^*)$ 
& $\alpha_e^{(2)}/10^3$ 
& $\rho_e/10^3$ 
\\
\hline
\vspace{-2mm}\\
\PDtabline{1}
\PDtabline{2}
\PDtabline{3}
\PDtabline{4}
\PDtabline{5}
\PDtabline{6}
\end{tabular}
\end{ruledtabular}
\end{table*}

\section{BBR clock shift}
Atomic units are employed throughout this and the following section. The usual definitions, $e=m_e=\hbar=4\pi\epsilon_0=1$, are supplemented with the additional definition $k_B=1$, where $k_B$ is Boltzmann's constant. We also define a reference temperature $T_0$ equivalent to 300 Kelvin and having a value $T_0=9.50\times10^{-4}$ in our system of units. 
We base our unit system on SI electromagnetic expressions, with the Bohr magneton being given by $\mu_B=e\hbar/2m_e=1/2$.
The speed of light $c$ is used in favor of the fine structure constant $\alpha$ in expressions to follow to avoid notational confusion; $c=\alpha^{-1}\approx137$ in atomic units.

The energy shift to clock state $n$ due to electric dipole coupling with thermal radiation reads
\begin{eqnarray}
\Delta E_n=-2\pi
\int_0^\infty \!\!u(\omega)\,\alpha_n(\omega)\,d\omega,
\label{Eq:dE}
\end{eqnarray}
where $u(\omega)$ is the spectral energy density at (angular) frequency $\omega$
and $\alpha_n(\omega)$ is the frequency-dependent polarizability,
\begin{eqnarray}
\alpha_n(\omega)=\frac{2}{3}\sum_{n^\prime\ne n}
\left|\langle n^\prime||\mathbf{D}||n\rangle\right|^2
\frac{\omega_{n^\prime n}}{\omega_{n^\prime n}^2-\omega^2}.
\label{Eq:pol}
\end{eqnarray}
The Cauchy principal value is implicitly taken for the integral of Eq.~(\ref{Eq:dE}), as well as for the integral of Eq.~(\ref{Eq:FW}) to follow \citep{FarWin81}.
For BBR, the spectral energy density $u(\omega)$ may be written in terms of temperature $T$ using Planck's law,
\begin{eqnarray*}
u(\omega)=\frac{1}{\pi^2c^3}\frac{\omega^3}{e^{\omega/T}-1},
\end{eqnarray*}
and it follows that the energy shift may be recast as
\begin{eqnarray}
\Delta E_n=-\frac{T^3}{c^3}\sum_{n^\prime\ne n}
\left|\langle n^\prime||\mathbf{D}||n\rangle\right|^2
F\left(\frac{\omega_{n^\prime n}}{T}\right),
\label{Eq:dE_FW}
\end{eqnarray}
where $F(y)$ is the function introduced by Farley and Wing \cite{FarWin81},
\begin{eqnarray}
F(y)=\frac{2}{3\pi}\int_0^\infty dx\left(\frac{1}{y+x}+\frac{1}{y-x}\right)\frac{x^3}{e^x-1}.
\label{Eq:FW}
\end{eqnarray}
The function $F(y)$ is displayed in Figure \ref{Fig:FW}.
The atomic transition frequencies all satisfy $\omega_{n^\prime n}\gg T_0$, and it follows that for room temperature (or below), the factor $F(\omega_{n^\prime n}/T)$ appearing in Eq.~(\ref{Eq:dE_FW}) may be well-approximated with the leading terms of the asymptotic expansion
$F(y)=4\pi^3/45y+32\pi^5/189y^3+32\pi^7/45y^5+\dots$ \cite{FarWin81,PorDer06}. 
The energy shift may then be decomposed into respective terms,
\begin{eqnarray}
\Delta E_n=
-\frac{2\pi^3T^4}{15c^3}\alpha_n^{(0)}
-\frac{16\pi^5T^6}{63c^3}\alpha_n^{(2)}
-\frac{16\pi^7T^8}{15c^3}\alpha_n^{(4)}
-\dots,
\label{Eq:dE3term}
\end{eqnarray}
where we have introduced the frequency-independent atomic factors $\alpha_n^{(k)}$,
\begin{eqnarray}
\alpha_n^{(k)}=\frac{2}{3}\sum_{n^\prime\ne n}
\frac{\left|\langle n^\prime||\mathbf{D}||n\rangle\right|^2}
{\omega_{n^\prime n}^{k+1}}.
\label{Eq:alphak}
\end{eqnarray}
Noting the equivalence between $\alpha_n^{(0)}$ and the static polarizability, $\alpha_n^{(0)}=\alpha_n(0)$, we identify the leading term in Eq.~(\ref{Eq:dE3term}) with the `static approximation' of the energy shift. 
To connect with Eq.~(1) of main text, we write $\Delta\nu_\mathrm{BBR}=\left(\Delta E_e-\Delta E_g\right)/2\pi$ with the static contribution factored out:
\begin{eqnarray}
\Delta\nu_\mathrm{BBR}=
-\frac{\pi^2T_0^4}{15c^3}\Delta\alpha^{(0)}\left(\frac{T}{T_0}\right)^4
\left[
1
+\frac{40\pi^2T_0^2}{21}\frac{\Delta\alpha^{(2)}}{\Delta\alpha^{(0)}}
\left(\frac{T}{T_0}\right)^2
+8\pi^4T_0^4\frac{\Delta\alpha^{(4)}}{\Delta\alpha^{(0)}}
\left(\frac{T}{T_0}\right)^4
+\dots
\right],
\label{Eq:deltanufortheguys}
\end{eqnarray}
where $\Delta\alpha^{(k)}\equiv\alpha_e^{(k)}-\alpha_g^{(k)}$ is the differential atomic factor taken between the excited ($e$) and ground ($g$) clock states. The correction factor $\eta_\mathrm{clock}(T)$ can then be equated to the terms in square brackets succeeding the leading 1.

\begin{figure}
\includegraphics[width=0.5\linewidth]{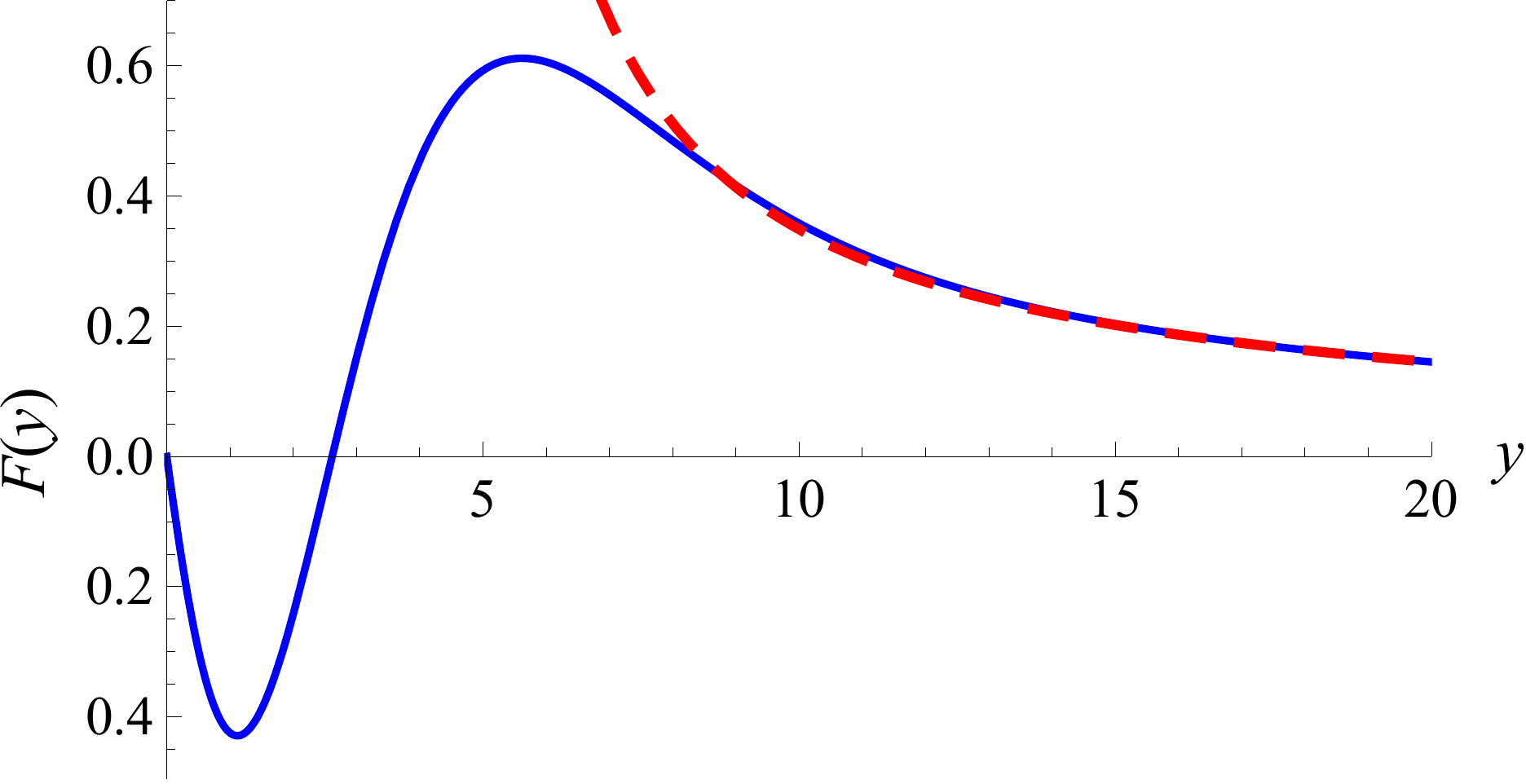}
\caption{(color online) Function $F(y)$ (solid) and it's asymptotic form truncated at the leading three terms  (dashed). For Yb, all transition frequencies satisfy $\omega_{n^\prime n}/T_0>34$. We find the three-term asymptotic expansion of $F(y)$ to be accurate to better than 4 parts in $10^5$ for $y>34$.}
\label{Fig:FW}
\end{figure}

\section{Evaluating $\eta_\mathrm{clock}(T)$}
With $\Delta\alpha^{(0)}$ known accurately from recent experimental measurement \cite{SheLemHin12etal}, the leading term of $\eta_\mathrm{clock}$ is determined by the differential atomic factor $\Delta\alpha^{(2)}$. Table \ref{Tab:alphas} tabulates contributions to $\Delta\alpha^{(2)}$, yielding a final result
\begin{eqnarray*}
\Delta\alpha^{(2)}=1.49(4)\times10^5.
\end{eqnarray*}
We see that the $\clockp\rightarrow\state{0}{1}$ transition dominates. While other contributions are smaller in comparison, they are nevertheless non-negligible at our accuracy.

As mentioned previously, theoretical uncertainty is difficult to assess for contributions from `mixed states' and `all others'. For this reason, we also consider an alternative extraction of $\Delta\alpha^{(2)}$ by defining a property
\begin{eqnarray*}
\rho_n&\equiv&\alpha_n^{(2)}+\frac{0.75}{\omega^{*2}}\left[\alpha_n(0)-\alpha_n(\omega^*)\right]
=\frac{2}{3}\sum_{n^\prime\ne n}
\left|\langle n^\prime||\mathbf{D}||n\rangle\right|^2
\frac{1}{\omega_{n^\prime n}^3}
\left(
1-0.75\frac{\omega_{n^\prime n}^2}{\omega_{n^\prime n}^2-\omega^{*2}}.
\right)
\end{eqnarray*}
From this definition, we get an expression for $\Delta\alpha^{(2)}$:
\begin{eqnarray*}
\Delta\alpha^{(2)}
=
\Delta\rho-\frac{0.75}{\omega^{*2}}
\left[\Delta\alpha(0)-\Delta\alpha(\omega^*)\right]
=\Delta\rho-30.3659(6)\times10^3,
\end{eqnarray*}
where we have utilized accurately known experimental parameters $\Delta\alpha(0)$, $\Delta\alpha(\omega^*)$, and $\omega^*$ in the last equality. Since the last term is known precisely, uncertainty in $\Delta\alpha^{(2)}$ is then commensurate with uncertainty in $\Delta\rho$.
Table \ref{Tab:alphas} gives contributions to $\Delta\rho$, and we find `mixed states' and `all others' to give negligible contribution in this case. With $\Delta\rho$ as given in Table \ref{Tab:alphas}, we then obtain
\begin{eqnarray*}
\Delta\alpha^{(2)}=1.48(5)\times10^5,
\end{eqnarray*}
which is in agreement with our result given above.

Using $\Delta\alpha^{(2)}=1.49(4)\times10^5$ we find the leading term to the dynamic correction factor to be:
\begin{eqnarray*}
\frac{40\pi^2T_0^2}{21}\frac{\Delta\alpha^{(2)}}{\Delta\alpha^{(0)}}
=0.0173(5)
\end{eqnarray*}
Contributions to $\alpha_n^{(k)}$ scale as $1/\omega_{n^\prime n}^{k+1}$. This explains why $\clockp\rightarrow\state{0}{1}$---whose transition frequency is less than half of all other transition frequencies---has an increased relative importance in $\Delta\alpha^{(2)}$ compared to $\Delta\alpha^{(0)}$. For $\Delta\alpha^{(4)}$, we find that $\clockp\rightarrow\state{0}{1}$ completely dominates, and we obtain for the next highest term of the dynamic correction factor:
\begin{eqnarray*}
8\pi^4T_0^4\frac{\Delta\alpha^{(4)}}{\Delta\alpha^{(0)}}
=0.00059(1).
\end{eqnarray*}
Higher-order terms are found to be negligible. Finally, putting these results together, we find
\begin{eqnarray*}
\eta_\mathrm{clock}(T)=0.0173(5)\left(\frac{T}{T_0}\right)^2+0.0006\left(\frac{T}{T_0}\right)^4,
\end{eqnarray*}
where uncertainty in the second term may be neglected.

\subsection{Magnetic dipole and higher-multipolar couplings}
Thus far we have confined our attention to dominant electric dipole ($E1$) coupling to the BBR field.
Additionally, the atom couples to the BBR field via magnetic dipole ($M1$) and higher multipolar ($E2$, $M2$, \dots) interactions. Porsev and Derevianko \cite{PorDer06} argued that for room temperature BBR, the $M1$ coupling could potentially lead to fractional frequency shifts on the level of $10^{-18}$ in optical lattice clocks. Higher multipolar couplings were shown to be suppressed below this level, and we neglect them here.

The $M1$ BBR shift is analogous to the $E1$ shift; it is given by Eq.~(\ref{Eq:dE}) with the substitution and $\alpha_n(\omega)\rightarrow\beta_n(\omega)/c^2$, with $\beta_n(\omega)$ being the frequency-dependent {\it magnetic} polarizability. $\beta_n(\omega)$ is defined analogously to $\alpha_n(\omega)$ [Eq.~(\ref{Eq:pol})], but with the magnetic dipole operator $\bm{\mu}$ replacing the electric dipole operator $\mathbf{D}$. The additional factor $c^2$ in the substitution accounts for the different magnitudes of electric and magnetic fields in the BBR spectrum.

Due to parity selection rules ($\bm{\mu}$ is an even-parity operator, whereas $\mathbf{D}$ is an odd-parity operator) the $M1$-allowed transitions differ from the $E1$-allowed transitions. In particular, the {\p} clock state has a low-frequency $M1$ transition to the neighboring {\pa} state of the same fine structure manifold. For this transition, $\omega_{n^\prime n}/T_0\approx3.4$, and it follows that the asymptotic expansion of $F(y)$ is not appropriate (see Fig~\ref{Fig:FW}). In this case, Eq.~(\ref{Eq:dE_FW}) (with $\mathbf{D}\rightarrow\bm{\mu}/c$) should be used directly. 

The $M1$ shift may be estimated by assuming the non-relativistic limit and absence of configuration mixing between states. In the non-relativistic limit, the magnetic dipole operator is given by the expression $\bm{\mu}=-\mu_B(\mathbf{L}+2\mathbf{S})$, where $\mathbf{L}$ and $\mathbf{S}$ are the total orbital and spin angular momenta of the electrons, respectively. In the absence of configuration mixing, it follows that the only non-vanishing $M1$ matrix element involving either clock state is the one connecting the {\p} clock state to the nearby {\pa} state. This matrix element may then be evaluated analytically, with the result $\left|\langle nsnp\,{\p}||\bm{\mu}||nsnp\,{\pa}\rangle\right|=\sqrt{2}\mu_B$. Within these approximations, the room temperature $M1$ BBR shift to the ${\p}$ clock level is found to be
\begin{eqnarray*}
\Delta E_{\p}^{(M1)}\approx
-\frac{T_0^3}{2c^5}\, F\left(\frac{\omega_\mathrm{fs}}{T_0}\right),
\end{eqnarray*}
where $\omega_\mathrm{fs}$ is the fine structure interval between the {\p} and {\pa} states, and the factor $F\left(\omega_\mathrm{fs}/T_0\right)\approx\!0.28$ for Yb. Interpreted as an additional contribution to $\eta_\mathrm{clock}(T_0)$, we find:
\begin{eqnarray*}
\eta_\mathrm{clock}^{(M1)}(T_0)\approx1\times10^{-5}.
\end{eqnarray*}
We therefore conclude that $M1$ coupling to the BBR field is negligible.

\newpage
\end{widetext}

\end{document}